\documentstyle[11pt]{article}
\input amssymb.sty
\begin{document}
\newcommand{\Si}{\Sigma}
\newcommand{\tr}{{\rm tr}}
\newcommand{\ad}{{\rm ad}}
\newcommand{\Ad}{{\rm Ad}}
\newcommand{\ti}[1]{\tilde{#1}}
\newcommand{\om}{\omega}
\newcommand{\Om}{\Omega}
\newcommand{\de}{\delta}
\newcommand{\al}{\alpha}
\newcommand{\te}{\theta}
\newcommand{\vth}{\vartheta}
\newcommand{\be}{\beta}
\newcommand{\la}{\lambda}
\newcommand{\La}{\Lambda}
\newcommand{\D}{\Delta}
\newcommand{\ve}{\varepsilon}
\newcommand{\ep}{\epsilon}
\newcommand{\vf}{\varphi}
\newcommand{\G}{\Gamma}
\newcommand{\ka}{\kappa}
\newcommand{\ip}{\hat{\upsilon}}
\newcommand{\Ip}{\hat{\Upsilon}}
\newcommand{\ga}{\gamma}
\newcommand{\ze}{\zeta}
\newcommand{\si}{\sigma}
\def\bfa{{\bf a}}
\def\bfb{{\bf b}}
\def\bfc{{\bf c}}
\def\bfd{{\bf d}}
\def\bfe{{\bf e}}
\def\bfm{{\bf m}}
\def\bfn{{\bf n}}
\def\bfp{{\bf p}}
\def\bfu{{\bf u}}
\def\bfv{{\bf v}}
\def\bft{{\bf t}}
\def\bfx{{\bf x}}
\def\bfg{{\bf g}}
\def\bfS{{\bf S}}
\def\bfJ{{\bf J}}
\newcommand{\li}{\lim_{n\rightarrow \infty}}
\newcommand{\mat}[4]{\left(\begin{array}{cc}{#1}&{#2}\\{#3}&{#4}
\end{array}\right)}
\newcommand{\thmat}[9]{\left(
\begin{array}{ccc}{#1}&{#2}&{#3}\\{#4}&{#5}&{#6}\\
{#7}&{#8}&{#9}
\end{array}\right)}
\newcommand{\beq}[1]{\begin{equation}\label{#1}}
\newcommand{\eq}{\end{equation}}
\newcommand{\beqn}[1]{\begin{eqnarray}\label{#1}}
\newcommand{\eqn}{\end{eqnarray}}
\newcommand{\p}{\partial}
\newcommand{\di}{{\rm diag}}
\newcommand{\oh}{\frac{1}{2}}
\newcommand{\su}{{\bf su_2}}
\newcommand{\uo}{{\bf u_1}}
\newcommand{\SL}{{\rm SL}(2,{\mathbb C})}
\newcommand{\GLN}{{\rm GL}(N,{\mathbb C})}
\def\slN{{\rm sl}(N, {\mathbb C})}
\def\SLN{{\rm SL}(N, {\mathbb C})}
\newcommand{\gln}{{\rm gl}(N, {\mathbb C})}
\newcommand{\PSL}{{\rm PSL}_2( {\mathbb Z})}
\def\f1#1{\frac{1}{#1}}
\def\lb{\lfloor}
\def\rb{\rfloor}
\def\sn{{\rm sn}}
\def\cn{{\rm cn}}
\def\dn{{\rm dn}}
\newcommand{\rar}{\rightarrow}
\newcommand{\upar}{\uparrow}
\newcommand{\sm}{\setminus}
\newcommand{\ms}{\mapsto}
\newcommand{\bp}{\bar{\partial}}
\newcommand{\bz}{\bar{z}}
\newcommand{\bA}{\bar{A}}
\newcommand{\bL}{\bar{L}}

\newcommand{\vtb}{\theta_{10}}
\newcommand{\vtc}{\theta_{00}}
\newcommand{\vtd}{\theta_{01}}

\newcommand{\sect}[1]{\setcounter{equation}{0}\section{#1}}
\renewcommand{\theequation}{\thesection.\arabic{equation}}
\newtheorem{predl}{Proposition}[section]
\newtheorem{defi}{Definition}[section]
\newtheorem{rem}{Remark}[section]
\newtheorem{cor}{Corollary}[section]
\newtheorem{lem}{Lemma}[section]
\newtheorem{theor}{Theorem}[section]
\newtheorem{stat}{Statement}[section]

\vspace{0.3in}
\begin{flushright}
 ITEP-TH-47/03\\
\end{flushright}
\vspace{10mm}
\begin{center}
{\Large{\bf {
Elliptic Linear Problem for Calogero-Inozemtsev Model and Painlev{\'e}
VI Equation
}}
}\\
\vspace{5mm}
A.Zotov \\
{\sf Institute for Theoretical and Experimental Physics, Moscow, Russia,}\\
{\em e-mail zotov@itep.ru}
\vspace{5mm}
\end{center}

\begin{abstract}
We introduce $3N\times 3N$ Lax pair with spectral parameter for
Calogero-Inozemtsev model. The one degree of freedom case appears
to have $2\times 2$ Lax representation. We derive it from the elliptic
Gaudin model via some reduction procedure and prove algebraic integrability.
This Lax pair provides elliptic
linear problem for the Painlev{\'e} VI equation in elliptic form.
\end{abstract}

\section{Introduction}
\setcounter{equation}{0}
\subsection{Lax pair for Calogero-Inozemtsev model}
The elliptic Calogero-Mosero model (CM) \cite{C}
provides a notable example of integrable many-body systems.
It is defined by its Hamiltonian
\beq{u1}
H^{CM}=\frac{1}{2}\sum\limits_{i=1}^N v_i^2+g^2\sum_{i>j}^N
\wp(u_i-u_j)
\eq
An important tool for investigating the integrable systems
is the Lax representation with a spectral parameter.
For CM model it was suggested by I.Krichever \cite{Kr0}.
An extension of Calogero type models
for root systems of simple Lie algebras was due to
M.Olshanetsky and A.Perelomov \cite{OP}.
Later the Lax pairs for these models were
constructed in \cite{DPh, BCS}.
The model of much current interest is the Calogero-Inozemtsev (CI) one
\cite{Inoz}. It is described by the
Hamiltonian
\beq{u2}
H^{CI}=\frac{1}{2}\sum\limits_{i=1}^N v_i^2+g^2\sum_{i>j}^N\left(
\wp(u_i-u_j)+\wp(u_i+u_j)\right)+\sum\limits_{i=1}^N\sum\limits_{\al=0}^3
\nu^2_\al\wp(u_i+\om_\al)
\eq
on an elliptic curve $\langle 1,\tau \rangle$,
$\om_\al=\{0, \frac{1}{2}, \frac{\tau}{2}, \frac{1+\tau}{2}\}$
with five free constants $g,\nu_\al$. It generalizes the $BC_N$ type CM
model.
In original paper by V.Inozemtsev
\cite{Inoz} the Lax representation was constructed and a principal
fact of existence of the spectral parameter was proved.
However, an explicit dependence on the spectral parameter failed to be found.
Following above mentioned results we suggest a new $3N\times 3N$ Lax
representation for CI model with explicit dependance on the spectral
parameter.

\subsection{Reduction from elliptic Gaudin model}
Another line of research is related to the Hitchin
approach \cite{H,LO,Kr} to the classical integrable systems.
The first concrete examples were constructed by N.Nekrasov \cite{N}.
The spin generalization of CM model \cite{KBBT} and the elliptic top
\cite{STSR} make up typical systems of this kind.
(A natural relationship between the top and the spin CM model
was found \cite{Has,LOZ}.)
However, spinless systems only of $A_N$ type were described
in the Hitchin-Nekrasov framework.
The problem is to find some reduction procedure which would freeze
the spin degrees of freedom in a way which produces root systems of
other types for spinless CM models. An example of this kind of reduction
is going to be introduced and applied to the $2\times 2$ elliptic
Gaudin model \cite{N} with four points on an elliptic curve (G4).
As a result we come to the one degree of freedom CI model described
by the Hamiltonian
\beq{u3}
H^{PCI}=\frac{1}{2}v^2+\sum\limits_{\al=0}^{3}\nu_\al^2\wp(u+\om_\al),
\eq
where "P" indicates its relation to Painlev{\'e} VI equation (see
below). Under the reduction we obtain $2\times 2$ Lax pair for this system
with spectral parameter on elliptic curve. Four constants $\nu_\al$
in (\ref{u3})
appear from the Casimirs of the orbits corresponding to
four marked points in G4 model. A particular case, when
$\nu_0=\nu_1=\nu_2=\nu_3$ transforms (\ref{u3}) to the
CM model with one degree of freedom.

\subsection{Spectral curve and algebraic integrability in
$2\times 2$ case}
We evaluate explicit expression for the spectral curve for G4 model
and find out that it is a 2-fold covering of ${\mathbb{CP}}^1$
branching at eight points. Its genus equals five.
The above mentioned reduction allows us to find a way
to decrease the genus from
five to one and thus provides the proof of the algebraic integrability
in $2\times 2$ case.

\subsection{Elliptic form of Painlev{\'e} VI equation}
The sixth Painlev{\'e} equation (PVI) \cite{Gam}
in the elliptic form \cite{Pain,Manin,BB}
is the nonautonomous version of equation of motion for (\ref{u3})
\beq{u4}
\frac{d^2u}{d\tau^2}=-\sum\limits_{\al=0}^{3}\nu_\al^2\wp'(u+\om_\al)
\eq
The standard form of PVI equation
 can be derived from the Schlesinger systems \cite{Sch} of the
isomonodromic deformations on ${{\mathbb{CP}}^1\backslash\{x_1,x_2,x_3,x_4\}}$
\cite{Fuchs,JM}.
It is desirable to have Schlesinger type description of PVI equation
on a torus.
A knowledge of $2\times 2$ Lax representation for (\ref{u3}),
is the key to solving this task.
We would like to notice that the isomonodromy preserving equations
on Riemann surfaces of arbitrary genus and in particular of genus one
has been much investigated in
\cite{Korotkin,LO2,Krich,Takas}.

Necessary elliptic function definitions and identities can be found
in the Appendix A.

\section{Lax Pair for Calogero-Inozemtsev Model}
\setcounter{equation}{0}
As it was mentioned in the Introduction
the CI model was defined by the following Hamiltonian:
\beq{b1}
H=\frac{1}{2}\sum\limits_{i=1}^N v_i^2+g^2\sum_{i>j}^N\left(
\wp(u_i-u_j)+\wp(u_i+u_j)\right)+\sum\limits_{i=1}^N\sum\limits_{\al=0}^3
\nu^2_\al\wp(u_i+\om_\al)
\eq
on an elliptic curve $\langle 1,\tau \rangle$, where
$\om_\al=\{0, \frac{1}{2}, \frac{\tau}{2}, \frac{1+\tau}{2}\}$
while $g,\nu_\al$ are arbitrary constants.

\begin{predl}
The Calogero-Inozemtsev model (\ref{b1}) admits $3N\times 3N$ Lax
representation with a spectral parameter
on the elliptic curve
\footnote{To have an appropriate sign behind the potential in (\ref{u2})
one should replace $g,\nu$ with $\sqrt{-1}g, \sqrt{-1}\nu$}
\beq{b2}
L=\thmat{V+A}{B_1}{-C_1}{B_2}{-V+A^T}{C_2}{-C_2}{C_1}{0},\ \ \ \
M=\thmat{D+A'}{B_1'}{-C_1'}{B_2'}{D+{A^T}'}{C_2'}{-C_2'}{C_1'}{D+E}
\eq
where all entries are $N\times N$ matrices. The
matrices $V, D, C_1$ and $C_2$ are diagonal while all others are offdiagonal:
\beq{b3}
\begin{array}{c}
A_{ij}=g(1-\delta_{ij})\Phi(z,u_i-u_j),\ \ \ \
E_{ij}=g(1-\delta_{ij})\left(\wp(u_i-u_j)-\wp(u_i+u_j)\right)
\\
V_{ij}=\delta_{ij}v_i,\ \ \ \
D_{ij}=g\delta_{ij}\sum\limits_{k\neq i}^N
\left(\wp(u_k-u_i)+\wp(u_k+u_i)\right),
\\
{C_1}_{ij}=\delta_{ij}\sum\limits_{\al=0}^3\nu_\al
\varphi_\al(z,\om_\al+u_i),\ \ \ \
{C_2}_{ij}=\delta_{ij}\sum\limits_{\al=0}^3\nu_\al
\varphi_\al(z,\om_\al-u_i),
\\
{B_1}_{ij}=g(1-\delta_{ij})\Phi(z,u_i+u_j),\ \ \ \
{B_2}_{ij}=g(1-\delta_{ij})\Phi(z,-u_i-u_j).
\end{array}
\eq
\end{predl}
It follows from the above proposition that there exists $3\times 3$
Lax representation which describes $PCI$ model (\ref{u3}).
Surprisingly, the following assertion holds:
\begin{predl}
The Painlev{\'e}-Calogero-Inozemtsev model (\ref{u3}) admits $2\times 2$ Lax
representation with a spectral parameter
on the elliptic curve
\beq{b4}
L^{PCI}=\mat{v}{0}{0}{-v}+\sum\limits_{\al=0}^3 L_\al^{PCI},\ \ \ \ \
L_\al^{PCI}
=\mat{0}{\nu_\al\vf_\al(z,\om_\al+u)}{\nu_\al\vf_\al(z,\om_\al-u)}{0}
\eq
$$
M^{PCI}=\sum\limits_{\al=0}^3 M_\al^{PCI},\ \ \ \ \
M_\al^{PCI}
=\mat{0}{\nu_\al\vf'_\al(z,\om_\al+u)}{\nu_\al\vf'_\al(z,\om_\al-u)}{0}
$$
\end{predl}
The existence of $2\times 2$ Lax pair (\ref{b4})
appears to be explicable on the
basis of its relation to ${sl(2,\mathbb{C})}$ the
elliptic Gaudin model \cite{N}
with four points on the elliptic curve (G4). It will be discussed
in the next section.

The proof of the above propositions is based on the identities given
in Appendix A. In particular from (\ref{ad2}) we easily come to
a useful equality:

\begin{lem}
For $\al\neq\be$ and matrices $L_\al^{PCI},M_\al^{PCI}$ (\ref{b4})
the following relation holds:
\beq{b5}
[L_\al^{PCI},M_\be^{PCI}]+[L_\be^{PCI},M_\al^{PCI}]=0
\eq
\end{lem}
{\em Proof}
\beq{b6}
\begin{array}{c}
\vf_\al(z,\om_\al+u)\vf'_\be(z,\om_\be-u)-
\vf_\al(z,\om_\al-u)\vf'_\be(z,\om_\be+u)+
\\
\vf_\be(z,\om_\be+u)\vf'_\al(z,\om_\al-u)-
\vf_\be(z,\om_\be-u)\vf'_\al(z,\om_\al+u)=
\\
\vf_{\al+\be}(z,\om_\al+\om_\be)(\wp(\om_\be-u)-\wp(\om_\al+u)+
\wp(\om_\al-u)-\wp(\om_\be+u))=0
\end{array}
\eq

\section{Algebraic Integrability in $2\times 2$ Case}
\setcounter{equation}{0}
\subsection{Elliptic Gaudin model and reduction to PCI model}
As indicated earlier, in the case of the single degree of freedom
the CI model is defined by the Hamiltonian
\beq{ar1}
H^{PCI}=\frac{1}{2}v^2+\sum\limits_{\al=0}^{3}\nu_\al^2\wp(u+\om_\al)
\eq
and equations of motion can be represented in the Lax form with
matrices (\ref{b4}).
The aim of the section is to prove the algebraic integrability
of the PCI model.
For this purpose let us consider the G4 model
\cite{N}. It describes $4$ degrees of freedom.
One corresponds to a the motion of pair of particles in the center of mass
frame while three others describe dynamics of a complicated manifold
$\{{\mathcal O}\times{\mathcal O}\times{\mathcal O}\times{\mathcal O}\}// T$,
where $T$ is the Cartan subgroup in $SL(2,{\mathbb{C}})$.
The Lax matrix of the G4 model is  $sl(2,{\mathbb{C}})$-valued function
on the torus with appropriate quasiperiodic properties and
four simple poles.
Residues of the four points are the orbits ${\mathcal O}_\al$ of the
coadjoint action by $SL(2,{\mathbb{C}})$.

Let us specify the Lax matrix for the G4 model on the doubled torus
$\langle 2,2\tau\rangle$:
\beq{ar2}
L^{G4}=\mat{v}{0}{0}{-v}+\sum\limits_{\al=0}^3 L_\al^{G4},\ \ \ \ \
L_\al^{G4}
=\mat{s_{11}^\al E_1(z-2\om_\al,2\tau)}
{\tilde{s}_{12}^{\al}\vf_\al(z,\om_\al+u)}
{\tilde{s}_{21}^\al\vf_\al(z,\om_\al-u)}
{-s_{11}^\al E_1(z-2\om_\al,2\tau)}
\eq
In doing so we imply functions to be defined on $\langle 1,\tau\rangle$
if the dependence on $2\tau$ is not given explicitly.
The four marked points are $\{2\om_\al\}$ or
$\{0,1,\tau,\tau+1\}$.
Notice that unlike the diagonal elements of residues
of (\ref{ar2}) $s_{11}^\al$ the offdiagonal
$\tilde{s}_{12}^\al$ and $\tilde{s}_{21}^\al$ do not correspond
to a certain orbit ${\mathcal O}_\al$ but to some linear combination
which can be expressed  through the use of the matrix $I$
(\ref{a33}):
\beq{a211}
\begin{array}{c}
s_{12}^{\rho}=\sum\limits_{\al=0}^3\bfe(-2u\p_\tau\om_{\rho})I_{\rho\al}
\tilde{s}_{12}^\al=
\sum\limits_{\al=0}^3
\bfe(2\om_\rho\p_\tau\om_\al-2(\om_\al+u)\p_\tau\om_\rho)\tilde{s}_{12}^{\al}
\\
s_{21}^{\rho}=\sum\limits_{\al=0}^3\bfe(2u\p_\tau\om_{\rho})I_{\rho\al}
\tilde{s}_{12}^\al=
\sum\limits_{\al=0}^3
\bfe(2\om_\rho\p_\tau\om_\al-2(\om_\al-u)\p_\tau\om_\rho)\tilde{s}_{21}^{\al}
\end{array}
\eq
The inverse change of variables comes from (\ref{a34}):
\beq{a212}
\tilde{s}_{12}^{\rho}=\frac{1}{4}\sum\limits_{\al=0}^3
\bfe(2u\p_\tau\om_{\al})I_{\rho\al}s_{12}^{\al},\ \ \ \
\tilde{s}_{21}^{\rho}=\frac{1}{4}\sum\limits_{\al=0}^3
\bfe(-2u\p_\tau\om_{\al})I_{\rho\al}s_{21}^{\al}
\eq

The spectral curve is defined by the equation
$$
\det(\lambda+L^{G4}(z))=0\ \ \ \ \
\hbox{or}\ \ \ \ \ \lambda^2+\det L^{G4}(z)=0
$$
Using (\ref{A.7a}) we have
\beq{ar3}
\begin{array}{c}
\lambda^2=
v^2+2v\sum\limits_{\al}s_{11}^\al E_1(z-2\om_\al,2\tau)+
\left(\sum\limits_{\al}s_{11}^\al E_1(z-2\om_\al,2\tau)\right)^2+
\\
\sum\limits_{\al}\tilde{s}_{12}^\al\tilde{s}_{21}^\al
\left(\wp(z)-\wp(u+\om_\al)\right)+
\sum\limits_{\al\neq\be}\tilde{s}_{12}^\al\tilde{s}_{21}^\be
\vf_\al(z,\om_\al+u)\vf_\be(z,\om_\be-u)
\end{array}
\eq
The Hamiltonian appears form the decomposition of function
$-\det L^{G4}(z)=\frac{1}{2}Tr\left(\left(L^{G4}(z)\right)^2\right)$:
\beq{ar4}
-\det L^{G4}(z)=\sum\limits_{\al=0}^3H_{2,\al}E_2(z-2\om_\al,2\tau)+
\sum\limits_{\al=0}^3H_{1,\al}E_1(z-2\om_\al,2\tau)+H_{2,0},
\eq
where $H_{2,\al}=C_{\al}$ are the Casimir functions of orbits
$s^\al$, $H_{1,\al}$ are the Hamiltonians linear in the momentum $v$ and
$H_{2,0}$ is the quadratic one.

The symmetry which underlies the reduction is generated by the
following involution:
\beq{ar5}
L^{G4}(z)\rightarrow -\sigma_1 L^{G4}(-z)\sigma_1,
\eq
where $\sigma_1=\mat{0}{1}{1}{0}$.
The reduction procedure implies that we should choose the eigenvalue
of the map (\ref{ar5}).
Choosing it to be $+1$ and keeping in mind
(\ref{P.2}) and (\ref{P.4}) we arrive to the following constraints:
\beq{ar6}
s_{11}^{\al}=0,\ \ \ \ \
\tilde{s}_{12}^{\al}=\tilde{s}_{21}^{\al}
\eq
It follows from (\ref{a212}) that
\beq{ar7}
\bfe(-2u\p_\tau\om_{\al})s_{21}^{\al}=
\bfe(2u\p_\tau\om_{\al})s_{12}^{\al}
\eq
If it is recalled that the Casimirs
$C_\al=\left(s_{11}^\al\right)^2+s_{12}^{\al}s_{21}^{\al}$
look like $C_\al=s_{12}^{\al}s_{21}^{\al}$ on shell (\ref{ar6}) we have
\beq{ar8}
s_{12}^{\al}=\bfe(-2u\p_\tau\om_{\al})\sqrt{C_\al},\ \ \ \
s_{21}^{\al}=\bfe(2u\p_\tau\om_{\al})\sqrt{C_\al}
\eq
And consequently from (\ref{a212})
\beq{ar9}
\tilde{s}_{12}^{\rho}=\tilde{s}_{21}^{\rho}=\frac{1}{4}\sum\limits_{\al=0}^3
I_{\rho\al}\sqrt{C_\al}=\nu_\rho.
\eq
Thereby the reduction provides transformation from the Lax matrix
of the Gaudin model (\ref{ar2}) to the one (\ref{b4}) of
Painlev{\'e}-Calogero-Inozemtsev.

The fact that the variables
$\tilde{s}_{12}^\al, \tilde{s}_{21}^\al$ and
$s_{11}^\al$ commute with the quadratic Hamiltonian with respect to
Poisson brackets on shell (\ref{ar6}) is discussed in Appendix B.

\subsection{Algebraic integrability}
To prove the algebraic integrability of PCI one should show
that the genus of its spectral curve equals $1$. The spectral
curve (\ref{ar3}) has genus $5$.
Indeed,  the r.h.s. of (\ref{ar3}) is the doubleperiodic function
on the elliptic curve $\langle
2,2\tau\rangle$ with second order poles at $4$ points.
Thus it has $8$ zeros and may be conceived  as a 2-fold covering
of the elliptic curve branching at eight points.
From the point of view of the spectral curve
the reduction by the involution (\ref{ar5}) implies
identification of points $z$ and $-z$.
The function $\wp(z)$ is even and has $8$ zeros on the torus $\langle
2,2\tau\rangle$. Thus it is natural to expect that
the unknown spectral curve could be written in terms of $\wp(z)$.

\begin{predl}
The spectral curve of the Painlev{\'e}-Calogero-Inozemtsev model
is of the form
\beq{ar10}
\lambda^2={\mathcal R}(X),\ \ X=\wp(z)
\eq
where ${\mathcal R}(X)$ is some rational function with four simple poles.
\end{predl}

Let us transform the last term from the r.h.s of
(\ref{ar3}). Using (\ref{ad4}) we have:

\beq{ar11}
\begin{array}{c}
\sum\limits_{\al\neq\be}\tilde{s}_{12}^\al\tilde{s}_{21}^\be
\vf_\al(z,\om_\al+u)\vf_\be(z,\om_\be-u)=
\\
\sum\limits_{\al\neq\be}\tilde{s}_{12}^\al\tilde{s}_{21}^\be
\vf_{\al+\be}(z,\om_\al+\om_\be)(E_1(z)+E_1(\om_\al+u)+E_1(\om_\be-u)-

E_1(z+\om_\al+\om_\be))=\\
=\sum\limits_{\al\neq\be}\tilde{s}_{12}^\al\tilde{s}_{21}^\be
\vf_{\al+\be}(z,\om_\al+\om_\be)(E_1(z)+E_1(\om_\al+\om_\be)-
E_1(z+\om_\al+\om_\be))+
\\
+\sum\limits_{\al\neq\be}\tilde{s}_{12}^\al\tilde{s}_{21}^\be
\vf_{\al+\be}(z,\om_\al+\om_\be)
(E_1(\om_\be-u)+E_1(\om_\al+u)-E_1(\om_\al+\om_\be))
\end{array}
\eq
The last sum vanishes under constraints (\ref{ar6})
\footnote{
$$
\begin{array}{c}
\sum\limits_{\al\neq\be}\nu^\al\nu^\be
\vf_{\al+\be}(z,\om_\al+\om_\be)
(E_1(\om_\be-u)+E_1(\om_\al+u)-E_1(\om_\al+\om_\be))=
\\
\frac{1}{2}\sum\limits_{\al\neq\be}\nu^\al\nu^\be
\vf_{\al+\be}(z,\om_\al+\om_\be)
(E_1(\om_\al-u)+E_1(\om_\al+u)
+E_1(\om_\be-u)+E_1(\om_\be+u)-2E_1(\om_\al)-2E_1(\om_\be))=0
\end{array}
$$}
and we come to the following equation for the spectral curve of the PCI
model:
\beq{ar12}
\begin{array}{c}
\lambda^2=v^2-\sum\limits_{\al}\nu_\al^2\wp(u+\om_\al)+
\\
+\wp(z)\sum\limits_{\al}\nu_\al^2+\sum\limits_{\al\neq\be}
\nu_\al\nu_\be\vf_{\al+\be}(z,\om_\al+\om_\be)
(E_1(z)+E_1(\om_\al+\om_\be)-E_1(z+\om_\al+\om_\be))
\end{array}
\eq
At this moment we need one more relation:
\beq{ar13}
\vf_{\al}(z,\om_\al)(E_1(z)+E_1(\om_\al)-
E_1(z+\om_\al))=\sum\limits_{\rho=0}^{3} I_{\al\rho}\wp(z-2\om_\rho,2\tau),
\eq
where matrix $I$ is defined in (\ref{a33}). The proof of (\ref{ar13})
is based on comparing the structure of singularities and (\ref{A7c}).
So we have
\beq{ar14}
\lambda^2=2H^{PCI}+\wp(z)\sum\limits_{\al}\nu_\al^2+
\sum\limits_{\al\neq\be}\nu_\al\nu_\be\sum\limits_{\rho}I_{\mu(\al,\be),\rho}
\wp(z-2\om_\rho,2\tau),
\eq
where the index $\mu(\al,\be)$ is uniquely defined from
\beq{ar15}
\om_{\mu(\al,\be)}=\om_\al+\om_\be\ \hbox{mod} (1,\tau)
\eq
Substituting
$\wp(z,\tau)=\sum\limits_{\al}\wp(z-2\om_\al,2\tau)$ into (\ref{ar14})
we come to the final answer:
\beq{ar16}
\lambda^2=2H^{PCI}+\sum\limits_{\rho=0}^3K_\rho\wp(z-2\om_\rho,2\tau),
\ \ \ \
K_{\rho}=\sum\limits_{\al,\be=0}^3\nu_\al\nu_\be I_{\mu(\al,\be),\rho}
\eq
To finish the proof we need one more step:
\beq{b55}
\wp(z+\om_\al)=\frac{1}{2}
\frac{\wp''(\om_\al)}{\wp(z)-\wp(\om_\al)}+\wp(\om_\al)
\eq
The r.h.s. of (\ref{ar16}) is the rational function with four simple
poles corresponding to $z=\{\om_\al\}$.
Thus the desirable curve of genus $1$ appears as the 2-fold covering
of ${\mathbb CP}^1$ branching at four points. In doing so we imply
$z$ as a coordinate on the torus
$\langle 1,\tau \rangle\supset \langle 2,2\tau \rangle\ $.

\section{Elliptic form of Painlev{\'e} VI equation}
As it was shown by Painlev{\'e} \cite{Pain} himself
the Painlev{\'e} VI
equation (PVI) can be represented in the following form
\beq{i1}
\frac{d^2u}{d\tau^2}=-\sum\limits_{\al=0}^{3}\nu_\al^2\wp'(u+\om_\al)
\eq
on the elliptic curve $\Sigma$ parameterized by $\langle 1,\tau \rangle$
where $\om_\al=\{0, \frac{1}{2}, \frac{\tau}{2}, \frac{1+\tau}{2}\}$
while $\nu_\al$ are arbitrary constants. Later the result was
rediscovered in \cite{Manin, BB}.
Initially B.Gambier \cite{Gam} found PVI in a more complicated
form
\beq{i2}
\frac{d^2X}{dt^2}=\frac{1}{2}\left(\frac{1}{X}+\frac{1}{X-1}+
\frac{1}{X-t}\right)\left(\frac{dX}{dt}\right)^2-
\left(\frac{1}{t}+\frac{1}{t-1}+\frac{1}{X-t}\right)\frac{dX}{dt}+
\eq
$$
+\frac{X(X-1)(X-t)}{t^2(t-1)^2}\left(\al+\be\frac{t}{X^2}+
\gamma\frac{t-1}{(X-1)^2}+\delta\frac{t(t-1)}{(X-t)^2}\right)
$$
which transforms into (\ref{i1}) by the following rules:
\beq{i3}
(u,\tau)\rightarrow\left(X=\frac{\wp(u)-\wp(\om_1)}{\wp(\om_2)-\wp(\om_1)},\
\tau=\frac{\wp(\om_3)-\wp(\om_1)}{\wp(\om_2)-\wp(\om_1)}
\right)
\eq
$$
(\nu_0^2, \nu_1^2, \nu_2^2, \nu_3^2)=
4\pi^2(\al,-\be,\frac{1}{2}\gamma,-\delta)
$$
The derivation of PVI equation as the preserving monodromy condition
was given by R.Fuchs \cite{Fuchs}. Namely it was shown that the PVI
equation can be derived from the Schlesinger system \cite{Sch, Okamoto}
\beq{i4}
\frac{\partial A_j}{\p \la_i}=\frac{[A_i,A_j]}{\la_i-\la_j},\ \ i\neq j,
\ \ \ \frac{\partial A_i}{\p \la_i}=-\sum\limits_{j\neq i}
\frac{[A_i,A_j]}{\la_i-\la_j}
\eq
on ${\mathbb{CP}}^1\backslash\{\lambda_1,\lambda_2,\lambda_3,\lambda_4\}$
where $A_j$ are ${\rm sl}(2, {\mathbb C})$-valued matrices.
It arises as the isomonodromy condition
\beq{i5}
\frac{\p \Psi}{\p \la_j}=-\frac{A_j}{\la-\la_j}\Psi
\eq
for a matrix-valued function $\Psi(\la)\in\SL$ which satisfies
the matrix differential equations
\beq{i6}
\frac{d\Psi}{d\la}=\sum\limits_{j=1}^4\frac{A_j}{\la-\la_j}\Psi
\eq
In other words the system of equations (\ref{i5}) and (\ref{i6})
is a linear problem for the Schlesinger equation (\ref{i4})
which appears to be equivalent to the rational form of PVI equation
(\ref{i2}). Here we come to a natural question. What is the analogue
of the Schlesinger equation on an elliptic curve
which leads to the
elliptic form of PVI (\ref{i1})?
In other words, we would like to find out if
there exists a pair of matrix-valued functions $L^{PVI}(z),M^{PVI}(z)$
on an elliptic curve $\Sigma$ which defines a linear problem
\beq{i7}
\left(\frac{\p}{\p z}+L^{PVI}(z)\right)\Psi(z)=0,\ \ \ \
\left(\frac{\p}{\p \tau}+M^{PVI}(z)\right)\Psi(z)=0
\eq
with the analogue of the Schlesinger equation
\beq{i8}
\frac{\p }{\p \tau}L^{PVI}-\frac{\p }{\p z}M^{PVI}=[L^{PVI},M^{PVI}]
\eq
to be equivalent to (\ref{i1}).

\begin{predl}
The Painlev{\'e} VI equation in elliptic form
is equivalent to
\beq{i18}
\p_\tau L^{PCI}-\p_z M^{PCI}=[L^{PCI},M^{PCI}]
\eq
with matrices $L^{PCI},M^{PCI}$ from (\ref{b4}).
\end{predl}

The proof is based on the  Proposition $2.2$ and identity:
\beq{i19}
\p_\tau \vf_\al(z,u+\om_\al)=\p_z\p_u\vf_\al(z,u+\om_\al)+\p_\tau u
\p_u\vf_\al(z,u+\om_\al)
\eq
For the case $\nu_0=\nu_1=\nu_2=\nu_3$ this statement was dicovered in
\cite{LO2}.

There is another way to describe the Painlev{\'e} VI equation which
is closed to the one considered in \cite{Korotkin, Takas}.
One may start from $sl(2,\mathbb{C})$ elliptic top \cite{STSR, LOZ}:
\beq{i20}
L^{top}(z)=\sum\limits_{\al=1}^3 S_\al\vf_\al(z,\al)\sigma_\al,\ \ \ \
M^{top}(z)=\sum\limits_{\al=1}^3 S_\al\vf'_\al(z,\al)\sigma_\al,
\eq
where $\sigma_\al$ are the Pauli matrices and $S_\al$ are the dynamical
variables. The Hamiltonian and Poisson brackets are:
\beq{i21}
H=\frac{1}{2}\sum\limits_{\al=1}^3 S_\al^2\wp(\om_\al),\ \ \
\{S_\al,S_\be\}=\varepsilon_{\al\be\gamma}S_\gamma
\eq
Now if we consider the top on the doubled torus $\langle 2,2\tau \rangle$
and put four orbits $S^{(\be)}$ into $0,1,\tau,\tau+1$ we obtain
the Lax matrix
\beq{i22}
L^{top4}=\sum\limits_{\al=1}^3 \tilde{S}^{(\be)}_\al\vf_\al(z,\al)\sigma_\al,
\ \ \ \
M^{top4}=\sum\limits_{\al=1}^3 \tilde{S}^{(\be)}_\al\vf'_\al(z,\al)\sigma_\al,
\eq
where as in the previous section
$$
\tilde{S}^{(\be)}=\frac{1}{4}\sum\limits_{\rho=0}^3 I_{\be\rho}S^{(\rho)}
$$
with matrix I defined in (\ref{a33}).
The Hamiltonian and Poisson brackets in this case are:
\beq{i23}
H=\frac{1}{32}\sum\limits_{\al=1}^3
\left(\sum\limits_{\be,\rho}I_{\be\rho}S^{(\rho)}_\al\right)^2\wp(\om_\al),
\ \ \
\{S_\al^{(\gamma)},S_\be^{(\rho)}\}=\delta^{\gamma \rho}
\varepsilon_{\al\be\gamma}S_\gamma
\eq
It follows from (\ref{c0})
that L-matrix defined in (\ref{i22}) is the doubleperiodic function on
the torus $\langle 2,2\tau \rangle$ and thus the sum of residues equals
zero
\beq{i24}
\sum\limits_{\rho=0}^3 S^{(\rho)}=0
\eq
The $SL(2,\mathbb{C})$ action saves these constrains. Consequently
the phase space of $top4$ model
looks like
${\mathcal O}_0\times{\mathcal O}_1\times{\mathcal O}_2
\times{\mathcal O}_3 //SL(2,\mathbb{C})$
and thus coincides with the one of the Schlesinger system (\ref{i4}).
The pair of matrices (\ref{i22}) obviously provides the Painlev{\'e} VI
equation in a sense of Proposition $4.1$.

\section{Acknowledgments}
The author is grateful to A.Levin and M.Olshanetsky for stimulating
discussions and to V.Pober-\\ ezhny and S.Oblezin for useful remarks.
The work was partially supported by grant INTAS 00-561, Grant
for Support of Scientific Schools NSh-1999.2003.2,
RFBR grant 03-02-17554, RFBR grant for support of young scientists
03-02-06453 and Federal Program No 40.052.1.1.1112.

\appendix
\section{Elliptic Functions}
\setcounter{equation}{0}
We summarize the main formulae for elliptic functions, borrowed
mainly from \cite{We} and \cite{Ma}.
We assume that $q=\exp 2\pi i\tau$, where $\tau$ is the modular parameter
of the elliptic curve $\langle 1,\tau \rangle$, half periods
$\om_\al=\{0, \frac{1}{2}, \frac{\tau}{2}, \frac{1+\tau}{2}\}$
and ${\bf e}(x)=\exp(2\pi\sqrt{-1}x)$.
We also deal with the doubled lattice $\langle 2,2\tau\rangle$ and
writing $f(z,w,2\tau)$ we mean that the function is defined on it
while writing $f(z,w)$ we imply that the function is defined on the
initial lattice $\langle 1,\tau \rangle$.

The basic element is the theta  function:
\beq{A.1a}
\vth(z|\tau)=q^{\frac
{1}{8}}\sum_{n\in {\bf Z}}(-1)^ne^{\pi i(n(n+1)\tau+2nz)}=
\eq
$$
q^{\frac{1}{8}}e^{-\frac{i\pi}{4}} (e^{i\pi z}-e^{-i\pi z})
\prod_{n=1}^\infty(1-q^n)(1-q^ne^{2i\pi z})(1-q^ne^{-2i\pi z})
 $$
{\sl The  Eisenstein functions}
\beq{A.1}
E_1(z|\tau)=\p_z\log\vth(z|\tau), ~~E_1(z|\tau)\sim\f1{z}-2\eta_1z,
\ \ \eta_1(\tau)=\ze(\frac{1}{2})
\eq

\beq{A.2}
E_2(z|\tau)=-\p_zE_1(z|\tau)=
\p_z^2\log\vth(z|\tau),
~~E_2(z|\tau)\sim\f1{z^2}+2\eta_1
\eq

{\sl Particular values}
\beq{A7a}
E_1(\om_\al)=-2\pi\sqrt{-1}\p_\tau\om_\al,
\eq
where $\p_{\tau}\om_\al=\{0, 0, \frac{1}{2}, \frac{1}{2}\}$
or equivalently
\beq{A7b}
E_1(\oh)=0,~~E_1(\frac{\tau}{2})=E_1(\frac{1+\tau}{2})=
-\pi\sqrt{-1}.
\eq
\beq{A7c}
\sum\limits_{\al=1}^3\wp(\om_\al)=0
\eq

The next important function is
\beq{A.3}
\phi(z,u)=
\frac
{\vth(u+z)\vth'(0)}
{\vth(u)\vth(z)}.
\eq
It has a pole at $z=0$ and
\beq{A.3a}
\phi(z,u)=\frac{1}{z}+E_1(u)+\frac{z}{2}(E_1^2(u)-\wp(u))+\ldots,
\eq
and
\beq{A3b}
\phi'(z,u)=\phi(z,u)(E_1(u+z)-E_1(u)).
\eq
We use prime for the derivation with respect to the second argument, i.e.
$$
\phi'(z,w)=\frac{\p}{\p y}\phi(z,y)|_{y=w}
$$

One of the most important notations is
\beq{a31}
\vf_\al(z,\om_\be+u)=\bfe(z\p_\tau\om_\al)\phi(z,\om_\be+u)
\eq

It should be mentioned that at $z=0$
\beq{a32}
\vf_\al(z,\om_\al)=\frac{1}{z}-\frac{z}{2}\wp(\om_\al)+\dots
\eq

{\sl Relations to the Weierstrass functions}
\beq{A.4}
\ze(z|\tau)=E_1(z|\tau)+2\eta_1(\tau)z,
\eq
\beq{A.5}
\wp(z|\tau)=E_2(z|\tau)-2\eta_1(\tau),
\eq
\beq{A.7}
\phi(z,u)=\exp(-2\eta_1uz)
\frac
{\si(u+z)}{\si(u)\si(z)}.
\eq
\beq{A.7a}
\phi(z,u)\phi(-u,z)=\wp(z)-\wp(u)=E_2(z)-E_2(u).
\eq

{\sl Parity}

\beq{P.1}
\vth(-z)=-\vth(z)
\eq
\beq{P.2}
E_1(-z)=-E_1(z)
\eq
\beq{P.3}
E_2(-z)=E_2(z)
\eq
\beq{P.4}
\phi(z,u)=\phi(z,u)=-\phi(-u,-z)
\eq

{\sl Behavior on the lattice}

\beq{A.11}
\vth(z+1)=-\te(z),~~~\vth(z+\tau)=-q^{-\oh}e^{-2\pi \sqrt{-1}z}\te(z),
\eq
\beq{A.12}
E_1(z+2\om_\al)=E_1(z)-4\pi \sqrt{-1}\p_\tau \om_\al:\ \ \
E_1(z+1)=E_1(z),~~~E_1(z+\tau)=E_1(z)-2\pi \sqrt{-1},
\eq
\beq{A.13}
E_2(z+2\om_\al)=E_2(z):\ \ \ E_2(z+1)=E_2(z),~~~E_2(z+\tau)=E_2(z),
\eq
\beq{A.14}
\phi(u+1,z)=\phi(z,u),~~~\phi(u+\tau,z)=e^{-2\pi \sqrt{-1}z}\phi(z,u).
\eq

\beq{c0}
\begin{array}{l}
\vf_\al(z,\om_\al+u)\ \ \ \ \ \ \ Res_{z=0}\ \ \ \ \ \ \ Res_{z={1}}
\ \ \ \ \ \ \ \ Res_{z=\tau}\ \ \ \ \ \ \ \ Res_{z={1+\tau}}
\\
\vf_0(z,\om_0+u)\ \ \ \ \ \ \ \ \ \ \ \ \ 1
\ \ \ \ \ \ \ \ \ \ \ \ \ \ \ \ 1\ \ \ \ \ \ \ \ \ \ \ \bfe(-u)
\ \ \ \ \ \ \ \ \ \ \ \bfe(-u)
\\
\vf_1(z,\om_1+u)\ \ \ \ \ \ \ \ \ \ \ \ \ 1
\ \ \ \ \ \ \ \ \ \ \ \ \ \ \ \ 1\ \ \ \ \ \ \ \ -\bfe(-u)
\ \ \ \ \ \ \ -\bfe(-u)
\\
\vf_2(z,\om_2+u)\ \ \ \ \ \ \ \ \ \ \ \ \ 1
\ \ \ \ \ \ \ \ \ \ \ \ \ -1\ \ \ \ \ \ \ \ \ \ \bfe(-u)
\ \ \ \ \ \ \ \ -\bfe(-u)
\\
\vf_3(z,\om_3+u)\ \ \ \ \ \ \ \ \ \ \ \ \ 1
\ \ \ \ \ \ \ \ \ \ \ \ \ -1\ \ \ \ \ \ \ -\bfe(-u)
\ \ \ \ \ \ \ \ \ \ \ \bfe(-u)
\end{array}
\eq

The above matrix is defined by
\beq{c01}
Res_{z=2\om_\rho}\vf_\al(z,\om_\al+u)=\bfe(2\om_\rho\p_\tau\om_\al-
2(\om_\al+u)\p_\tau\om_\rho)
\eq

For the symmetric $4\times 4$ matrix $Res_{z=2\om_\rho}\vf_\al(z,\om_\al)$
we keep notation
\beq{a33}
I_{\rho\al}=Res_{z=2\om_\rho}\vf_\al(z,\om_\al)=
\bfe(2\om_\rho\p_\tau\om_\al-2\om_\al\p_\tau\om_\rho)
\eq
or
$$
I=\left(
\begin{array}{l}
1\ \ \ \ 1\ \ \ \ 1\ \ \ \ 1
\\
1\ \ \ \ 1\ -1\ -1
\\
1\ -1\ \ \ \ 1\ -1
\\
1\ -1\ -1\ \ \ \ 1
\end{array}\right)
$$
Note that
\beq{a34}
{I^{-1}}_{\al\be}=\frac{1}{4}I_{\al\be}
\eq
{\sl Addition formula}

\beq{ad1}
\phi(z,u)\p_v\phi(z,v)-\phi(z,v)\p_u\phi(z,u)=(E_2(v)-E_2(u))\phi(z,u+v),
\eq
or
\beq{ad2}
\phi(z,u)\p_v\phi(z,v)-\phi(z,v)\p_u\phi(z,u)=(\wp(v)-\wp(u))\phi(z,u+v).
\eq

In fact, $\phi(z,u)$ satisfies more general relation which follows from the
Fay three-section formula
\beq{ad3}
\phi(u_1,z_1)\phi(u_2,z_2)-\phi(u_1+u_2,z_1)\phi(u_2,z_2-z_1)-
\phi(u_1+u_2,z_2)\phi(u_1,z_1-z_2)=0
\eq
A particular case of this formula is
\beq{ad4}
\phi(u_1,z)\phi(u_2,z)=\phi(u_1+u_2,z)(E_1(u_1)+E_1(u_2)+E_1(z)-
E_1(u_1+u_2+z))
\eq

\section{Comments on Reduction Procedure}
\setcounter{equation}{0}
Let us have a look at the generating function of the Hamiltonians:
\beq{a10}
\begin{array}{c}
\frac{1}{2}Tr\left(L^{G4}(z)\right)^2=v^2+
2v\sum\limits_{\al}s_{11}^\al E_1(z-2\om_\al,2\tau)+
\left(\sum\limits_{\al}s_{11}^\al E_1(z-2\om_\al,2\tau)\right)^2+
\\
\sum\limits_{\al}\tilde{s}_{12}^\al\tilde{s}_{21}^\al
\left(\wp(z)-\wp(u+\om_\al)\right)+
\sum\limits_{\al\neq\be}\tilde{s}_{12}^\al\tilde{s}_{21}^\be
\vf_\al(z,\om_\al+u)\vf_\be(z,\om_\be-u)
\end{array}
\eq
Consider infinitesimal deformation from the constraints
$$
s_{11}^{\al}=0,\ \ \ \ \
\tilde{s}_{12}^{\al}=\tilde{s}_{21}^{\al}
$$
in the form
$$
\delta s_{11}^\al=\epsilon^\al
$$
Since the change of variables $s\rightarrow \tilde{s}$ is linear
we have
\beq{a11}
\begin{array}{c}
\tilde{s}_{12}^\al=\nu^\al+\tilde{\epsilon}^\al
\\
\tilde{s}_{21}^\al=\nu^\al-\tilde{\epsilon}^\al,
\end{array}
\eq
where $\tilde{\epsilon}$ are linear in $\epsilon$.
We would like to show that the deformation of the quadratic Hamiltonian
$H_{2,0}$ from (\ref{ar4}) is quadratic in $\epsilon$ on shell.

Substituting the deformations into (\ref{a10}) we find that
all terms obviously satisfy our assumption except the last one:
\beq{a12}
\begin{array}{c}
\delta\sum\limits_{\al\neq\be}\tilde{s}_{12}^\al\tilde{s}_{21}^\be
\vf_\al(z,\om_\al+u)\vf_\be(z,\om_\be-u)=
\\
\delta\sum\limits_{\al\neq\be}\tilde{s}_{12}^\al\tilde{s}_{21}^\be
\vf_{\al+\be}(z,\om_\al+\om_\be)(E_1(z)+E_1(\om_\al+u)+E_1(\om_\be-u)
\\
-E_1(z+\om_\al+\om_\be))
=\sum\limits_{\al\neq\be}(\tilde{\epsilon}^\al\nu^\be-
\tilde{\epsilon}^\be\nu^\al)
\vf_{\al+\be}(z,\om_\al+\om_\be)(E_1(z)+E_1(\om_\al+u)
\\
+E_1(\om_\be-u)-E_1(z+\om_\al+\om_\be))=
\\
\sum\limits_{\al\neq\be}(\tilde{\epsilon}^\al\nu^\be-
\tilde{\epsilon}^\be\nu^\al)
\vf_{\al+\be}(z,\om_\al+\om_\be)(E_1(\om_\al+u)
+E_1(\om_\be-u))
\end{array}
\eq
However this expression does not provide dynamics via quadratic Hamiltonian
since
\beq{a13}
\vf_\al(z,\om_\al)\sim \frac{1}{z}-\frac{z}{2}\wp(\om_\al)
\eq

\bigskip

\small{

}

\end{document}